\documentstyle[prl,aps,twocolumn,epsfig]{revtex}
\begin{document} 
\preprint{} 
\draft 

\title{ Depletion of the Dark Soliton:              \\
        the Anomalous Mode of the Bogoliubov theory   }

\author{ Jacek Dziarmaga$^{1,2}$,
         and Krzysztof Sacha$^1$}

\address{
${}^1$ Intytut Fizyki Uniwersytetu Jagiello\'nskiego,
       Reymonta 4, 30-059 Krak\'ow, Poland \\
${}^2$ Los Alamos National Laboratory,
       Theory Division T-6, Los Alamos, 
       New Mexico 87545, USA
}

\maketitle 

\begin{abstract}
Quantum depletion from an atomic quasi one dimensional Bose-Einstein
condensate with a dark soliton is studied in a framework of the Bogoliubov
theory. Depletion is dominated by an anomalous mode localized in a notch
of the condensate wave function. Depletion in the anomalous mode requires
different treatment than depletion without anomalous modes. In particular,
quantum depletion in the Bogoliubov vacuum of the anomalous mode is
experimentally irrelevant. A dark soliton is initially prepared in a state
with minimal depletion which is not a stationary state of the Bogoliubov
theory. The notch fills up with incoherent atoms depleted from the
condensate. For realistic parameters the filling time can be as short as
10 ms.
\end{abstract}


\section{ Introduction }

   Since first experimental realizations of atomic Bose-Einstein
condensates (BEC) \cite{nobel} the condensates have been subject to
intensive experimental and theoretical research. The condensates are
relatively easy to manipulate and a rich variety of phenomena can be
studied in a well controlled way. One example is a dark soliton in
analogy to nonlinear optics \cite{dark,Fedichev,Anglin}. The dark soliton
is a place in a quasi one dimensional condensate where a condensate wave
function $\phi(x)$ changes its sign. Its wave function has a notch where
density of the condensate is zero. This notch appears as a dark spot on
a bright cigar shaped image of a condensate. Dark solitons were realized
in the experiment of the Hannover group \cite{Hannover} and their 3
dimensional analogs were realized in Ref.~\cite{3D}.

  A wave function of the dark soliton is proportional to

\begin{equation}
\tanh[(x-q)/l] \;,
\label{tanh}
\end{equation}
where $l$ is a local value of the healing length at the position $q$ of
the soliton. The density of coherent atoms in the notch at $x=q$ is zero.
This creates a hole that might be filled up with incoherent atoms. Given
that interactions between atoms are repulsive, depletion of atoms from the
condensate to the notch is energetically favourable.

  Condensate with a soliton at $q=0$ has an antisymmetric wavefunction of
$x$. This means that all $N$ atoms are in the same antisymmetric single
particle state. The antisymmetry of the condensate wavefunction is
preserved by evolution according to the Gross-Pitaevskii 
equation. However, the full
$N$-body Hamiltonian depletes pairs of atoms from the antisymmetric
condensate wavefunction to symmetric modes of the trap. This two atom
process does not change the symmetry of the $N$-body state. The symmetric
modes have nonzero density in the soliton notch at $x=0$. Simply because the
condensate density is zero at $x=0$, the $x=0$ is where we can clearly see the
incoherent atoms. In our previous paper \cite{dks} we studied a toy model
of the dark soliton. We truncated single particle Hilbert space to 3
lowest modes of a one dimensional harmonic trap. This is a self-consistent
approximation for small $N$. We have shown that a state where all atoms
are condensed in the same antisymmetric wave function is not a stationary
state of a quantum Hamiltonian. Atoms are gradually depleted from the
antisymmetric condensate wave function and fill its notch. 
In this
paper we reach similar conclusions in the opposite regime of very large
$N$ which is relevant for the Hannover experiment \cite{Hannover}. We use
as a tool the Bogoliubov theory with noninteracting quasiparticles.

  In the Bogoliubov theory one considers small fluctuations around a
stationary condensate background. In a uniform condensate the soliton
(\ref{tanh}) has a translational mode $\partial_q\tanh[(x-q)/l]$ with zero
frequency. The same soliton in an nonuniform condensate in a harmonic trap
has an anomalous mode with negative frequency equal to $-1/\sqrt{2}$ times
the trap frequency $\omega_x$ \cite{Anglin}. The nonuniformity breaks the
translational invariance, moving the soliton away from the maximal density
in the center of the trap lowers its energy. We will see that the
anomalous mode is responsible for depletion of atoms from a condensate to
the soliton notch. Depletion in the Bogoliubov vacuum has been recently
studied in Ref.\cite{Law}. However, we argue that depletion in the
Bogoliubov vacuum analyzed in Ref.~\cite{Law} is not experimentally
relevant. We construct a quantum state with minimal depletion which
approximates the state of the system right after creation of the dark
soliton by phase imprinting \cite{Hannover}.  Time evolution of such a
state is responsible for greying of the dark soliton.

  In Ref.~\cite{Fedichev} it is suggested that deviations from the mean
field description of the dark soliton observed experimentally
\cite{Hannover} can be explained by dissipation due to collisions of the
condensate with a thermal cloud of noncondensed atoms. There is no doubt
that dissipation influences dynamics of the soliton in that experiment.
However, we show that the dark soliton is greying even in the absence of
any thermal cloud or even in 1D where the cloud decouples from the
soliton \cite{Fedichev}. 
The notch fills up with incoherent atoms quantum depleted from a
condensate. Our calculations, which are idealized with respect to the
experiment, give a greying time of ${\cal O}(10{\rm ms})$, which is
consistent with the experimental observation.

  This paper is organized as follows. Section II gives several definitions
and approximate formulas valid in the Thomas-Fermi regime of strong
interactions. Section III gives Thomas-Fermi approximate analytic
expressions for the frequency and the eigenfunctions of the anomalous
mode. In Section IV we work out a general expression for a time-dependent
number of atoms depleted in the anomalous mode. In Section V the number of
depleted atoms is calculated in the Bogoliubov vacuum of the anomalous
mode. In Section VI we construct a quantum state with minimal depletion
which approximates the state of the system right after creation of the
dark soliton. In Section VII we follow time evolution of the number and
density of depleted atoms, and estimate the greying time after which the
notch fills up with incoherent atoms. Section VIII contains discussion and
conclusions.

\section{ The dark soliton }

  In dimensionless oscillator units a Gross-Pitaevskii equation
\cite{GPE} for the condensate wave function $\phi(t,x)$ in a harmonic trap
is

\begin{equation} 
i \partial_t \phi = 
- \frac12 \partial_x^2 \phi + 
\frac12 x^2 \phi + 
g |\phi|^2 \phi \;. 
\label{GP} 
\end{equation}
$\phi(t,x)$ is normalized to $1$. The ground state of this equation is a
stationary state $\phi_0(x)\exp(-i\mu_0 t)$ with a symmetric $\phi_0$,
$\phi_0(x)=\phi_0(-x)$. In the Thomas-Fermi regime, $g\gg 1\;$, $\phi_0$
can be approximated by

\begin{equation}
\phi_0(x) \;
\stackrel{g\gg 1}{\cong} \; 
\sqrt{ \frac{2\mu_0-x^2}{2g} }\;.
\label{phi0}
\end{equation}
This wave function is normalized to $1$ provided that $\mu_0=(3g/2)^{2/3}/2$.
It is localized and varies on a lengthscale of

\begin{equation}
R \; = \; \sqrt{2\mu_0} \;=\; \left(\frac{3g}{2}\right)^{1/3} \;,
\label{R}
\end{equation}
the so called Thomas-Fermi radius.

 The lowest antisymmetric stationary state of the Gross-Pitaevskii
equation $\phi_1(x)\exp(-i\mu_1 t)$ is a dark soliton localized in the
center of the trap. In the Thomas-Fermi regime $\phi_1$ can be
approximated by the ansatz

\begin{equation}
\phi_1(x) \; \stackrel{g\gg 1}{\cong} \;
\phi_0(x) \; \tanh\left(x/l_0\right) \;.
\label{phi1}
\end{equation}
Here

\begin{equation}
l_0 \;= \; 
\frac{1}{g^{1/2}\phi_0(0)} \;=\; 
\sqrt{2}\left(\frac{2}{3g}\right)^{1/3} \;
\label{l0}
\end{equation}
is a healing length of the condensate at the condensate peak density in
the center of the trap. The soliton solution in Eq.(\ref{phi1}) has a
notch at $x=0$. We note that with increasing $g$ the Thomas-Fermi radius
$R$ is growing while the healing length $l_0$ tends to zero. For
$g\gg 1$ the width of the soliton is very small as compared to the size
of the condensate, $l_0\ll R$.

\section{ The anomalous mode }

  Bogoliubov equations \cite{GPE,castin} for small perturbations on the
dark soliton background are

\begin{eqnarray}
&&
-\frac12 u^{''} + 
\frac12 x^2 u +
2g\phi_1^2 u -
\mu_1 u +
g\phi_1^2 v \;=\;
+\omega u \;,
\nonumber\\
&&
-\frac12 v^{''} +
\frac12 x^2 v +
2g\phi_1^2 v -
\mu_1 v +
g\phi_1^2 u \;=\;
-\omega v \;.
\label{uv}
\end{eqnarray}
Here $^{'}=d/dx$. We introduce two functions $f_{\pm}(x)=u(x)\pm v(x)$ and
combine Eqs.(\ref{uv}) into

\begin{eqnarray}
&&
-\frac12 f^{''}_+ +
\frac12 x^2 f_+ +
3g\phi_1^2 f_+ -
\mu_1 f_+ \;=\;
\omega f_- \;,
\label{p}\\
&&
-\frac12 f^{''}_- +
\frac12 x^2 f_- +
g\phi_1^2 f_- -
\mu_1 f_- \;=\;
\omega f_+ \;.
\label{m}
\end{eqnarray}
Eqs.(\ref{p},\ref{m}) have a solution with negative $\omega$
\cite{Fedichev,Anglin}, the so called anomalous mode. In the Thomas-Fermi
regime $\omega$ can be approximated by \cite{Fedichev,Anglin}

\begin{equation}
\omega\;\stackrel{g\gg 1}{\cong}\; \frac{-1}{\sqrt{2}} \;,
\label{omega}
\end{equation}
i.e. $-1/\sqrt{2}$ times the frequency of the harmonic trap. This
Thomas-Fermi value of $\omega$ follows from effective equations of motion
for a position of a dark soliton in Ref.\cite{Anglin}, and from a
perturbative $(g\gg 1)$ treatment of the Bogoliubov equations
\cite{Fedichev}.

  The anomalous mode is a perturbative approximation to the motion of the
dark soliton with respect to the background condensate. In
the limit of $g\to\infty$ the Thomas-Fermi radius of the condensate
(\ref{R}) is growing to infinity and at the same time the healing length
(\ref{l0}) is shrinking to $0$. In this limit the condensate is
uniform on the lengthscale of the soliton width and the anomalous mode of
Eqs.(\ref{p},\ref{m}) becomes

\begin{eqnarray}
&&
\lim_{g\to\infty} f_+(x) \;\propto\;
\frac{1}{\cosh^2\left(x/l_0\right)} \;,
\nonumber\\
&&
\lim_{g\to\infty} f_-(x) \;=\; 0 \;.
\label{limit}
\end{eqnarray}
This $f_+(x)$ is simply a translational mode of the soliton in an almost
uniform condensate. This limit gives us the leading $g\gg 1$ term in the
expression for $f_+(x)$.

   We use Eq.(\ref{m}) to find the leading term in $f_-(x)$. To begin with
we note that $\phi_0(x)$ satisfies

\begin{equation}
-\frac12 \phi_0^{''} +
\frac12 x^2 \phi_0 +
g \phi_0^3 -
\mu_0 \phi_0 \;=\;0\;.
\label{phi0eq}
\end{equation}
Comparing Eq.(\ref{phi0eq}) with Eq.(\ref{m}) one may conclude that
$f_-(x)\cong \alpha\phi_0(x)$ with a certain constant $\alpha$.
Substitution of this ansatz and of leading $f_+(x)$ from Eqs.(\ref{limit})
into Eq.(\ref{m}) and then using Eq.(\ref{phi0eq}) gives an equation

\begin{eqnarray}
&& -(\mu_1-\mu_0)\alpha\phi_0(x) \;+\; 
\nonumber\\
&& g\left[ \phi_1^2(x)-\phi_0^2(x) \right]\alpha\phi_0(x)\;=\;
   \frac{\omega}{\cosh^2\left(x/l_0\right)} \;.
\nonumber
\end{eqnarray}
We expect that with a right choice of $\alpha$ this equation is approximately
satisfied for $g\gg 1$. As $\mu_1-\mu_0={\cal O}(g^0)$ the first term on
the left hand side is small as compared to the second term and can be
neglected. The second term is localized on a healing length around $x=0$
and the slow factor $\phi_0(x)$ in this term can be replaced by $\phi_0(0)$.
Finally we can use the $g\gg 1$ formulas in Eqs.(\ref{phi0},\ref{phi1}) to
get

\begin{equation}
g\left[ \tanh^2\left(x/l_0\right) - 1 \right]
\alpha\phi_0^3(0)\;=\;
\frac{\omega}{\cosh^2\left(x/l_0\right)} \;.
\end{equation} 
This equation is satisfied by 

\begin{equation}
\alpha \;=\;
\frac{-\omega}{g\phi_0^3(0)} \;=\;
\frac{4}{3\sqrt{g}}\;,
\end{equation}
which defines the leading term in $f_-(x)$. After normalization, so that
$\int dx\;(|u|^2-|v|^2)\equiv\int dx\;f_+f_-=1$, we get

\begin{eqnarray}
&&
f_+(x)\; \stackrel{g\gg 1}{\cong}\;
\frac{\sqrt{3g}}{2\sqrt{2}\;\cosh^2\left(x/l_0\right)}
\label{fp}\\
&&
f_-(x)\; \stackrel{g\gg 1}{\cong}\;
\sqrt{\frac{2}{3}}\;\phi_0(x)\;.
\label{fm}
\end{eqnarray}
We note that $f_+(x)$ is localized within the notch of the dark soliton
(\ref{phi1}).

\section{ Number of depleted atoms }

  We will argue that depletion of atoms from the condensate fills up the
notch in the condensate wave function (\ref{phi1}) with incoherent atoms.
This effect will appear experimentally as greying of the dark soliton.
Solution of the equations (\ref{uv}) reveals many different modes
corresponding to the Bogoliubov spectrum. However, according to
the numerical results in Ref.\cite{Law}, only the anomalous mode
has a wave function localized in the soliton notch \cite{Law}, compare
Eq.(\ref{fp}). From all the modes the anomalous mode will contribute the
most to the density of incoherent atoms filling the notch. This expectation
is confirmed by numerical calculations of incoherent atoms density
distribution in the Bogoliubov vacuum state \cite{Law}, where the
density, which is strongly peaked in the notch, is dominated by a
contribution from the anomalous mode. In the following we truncate the
Bogoliubov spectrum to the anomalous mode alone.

  Having constructed $f_\pm=u\pm v$ we can calculate an operator
corresponding to a number of depleted atoms in the anomalous mode

\begin{equation}
d\hat N(t)\;=\;\int dx\;\delta\hat\phi^\dagger(x,t)\;\delta\hat\phi(x,t)\;,
\end{equation}
where
\begin{equation}
\delta\hat\phi(x,t)\;=\;
u(x)\;   e^{-i\omega t}\;\hat a\;+\;
v^*(x)\; e^{+i\omega t}\;\hat a^\dagger
\label{deltaphi}
\end{equation}
is an annihilation operator of an incoherent atom in the anomalous mode.
The operators $\hat a$, $\hat a^\dagger$ annihilate/create a Bogoliubov
quasiparticle and they fulfill the usual bosonic commutation relation,
$[a,a^{\dagger}]=1$. We use Heisenberg picture defined by the Bogoliubov
Hamiltonian of the anomalous mode

\begin{equation}
H_B\;=\;-|\omega| \hat a^{\dagger} \hat a \;.
\label{HB}
\end{equation}
The operator of the number of depleted atoms is

\begin{eqnarray}
d\hat N(t)\;=\;\int dx\;
&[& |u|^2 \hat a^{\dagger} \hat a +
    |v|^2 \hat a \hat a^{\dagger} +
\nonumber\\
&& uv e^{-2i\omega t} \hat a \hat a +
   u^{*}v^{*} e^{+2i\omega t} \hat a^{\dagger} \hat a^{\dagger} ]\;.
\label{dN}
\end{eqnarray}
The density of incoherent atoms is the expectation value of the integrand.

\section{ Depletion in the Bogoliubov vacuum state }

  For a condensate in a ground state like in Eq.(\ref{phi0}) the
Bogoliubov
spectrum has only positive frequencies (i.e. there are no anomalous
modes). The state with no quasiparticles, the Bogoliubov vacuum, is the
ground state of the Bogoliubov Hamiltonian. In the absence of any
anomalous
modes it is natural to calculate the depletion in the Bogoliubov vacuum --
the $T=0$ state of thermodynamic equilibrium \cite{castin}.

  Following the numerical calculation in Ref.\cite{Law} we calculate
analytically for $g\gg 1$ a number of depleted atoms in the vacuum state
of the anomalous mode i.e. the state $|0\rangle_a$ annihilated by $\hat
a$, $\hat a|0\rangle_a=0$. This vacuum state is an eigenstate of the
Bogoliubov Hamiltonian (\ref{HB}) so the number is stationary

\begin{equation}
dN_a\;=\;_a\langle 0|d\hat N(t)|0\rangle_a\;=\;
\int dx\;|v(x)|^2\;,
\end{equation}
and scales with $g$ as $dN_a\propto g^{2/3}$. We compare the density of
depleted atoms in the soliton notch

\begin{equation}
\frac{dN_a}{l_0}\;\propto\;g\;\propto\;N\;,
\end{equation}
(where $N$ is the total number of atoms in the system) to the density of
atoms in the condensate

\begin{equation}
\frac{N}{R}\;\propto\;\frac{N}{g^{1/3}}\;\propto\;N^{2/3}\;.
\label{dbec}
\end{equation}
We see that, for a sufficiently large total number of atoms $N$ the
density of incoherent atoms in the notch will exceed the density of the
background condensate. The notch will not be visible. When the density of
depleted atoms is comparable to the density of the condensate one cannot
neglect interactions between quasiparticles. However, because of the
potential instability which shows up in the negative frequency of the
anomalous mode, we qualitatively expect that, unlike in more common
situations where all Bogolubov frequencies are positive, collisions
between quasiparticles will not suppress depletion.

  In Fig.1 we show condensate and total density plots in the Bogolubov
vacuum for the parameters of the Hannover experiment \cite{Hannover}
($N=1.5\cdot 10^5$ $^{87}$Rb atoms in the harmonic trap with
$\omega_x=2\pi\cdot 14\;$s$^{-1}$ and $\omega_{\perp}=2\pi\cdot
425\;$s$^{-1}$; $g=7500$ \cite{g1d}). As we can see, even for very
reasonable trap parameters and moderate $N$ the soliton notch is
substantially filled with incoherent atoms. For a larger $N$, the soliton
would not be observed at all.

\begin{figure}
%\centering \leavevmode \epsfxsize=4cm
\centering
{\epsfig{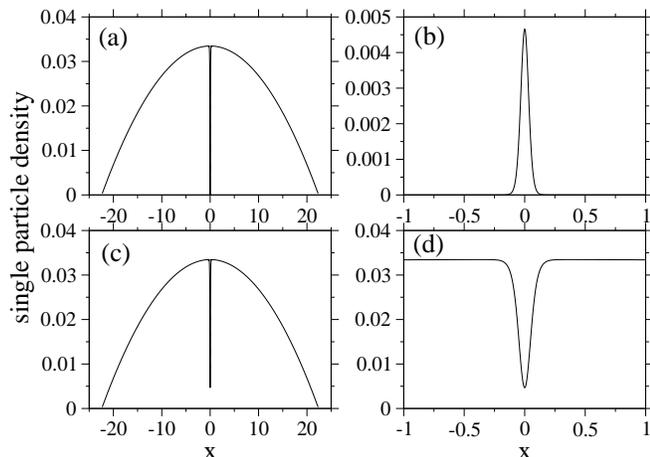}}
\caption{ Single particle density function corresponding to the experimental
parameters {\protect \cite{Hannover}}. The coherent part $|\phi_1(x)|^2$
[Eq.~({\protect \ref{phi1}})] is depicted in panel~(a), the incoherent part
$|v(x)|^2$ corresponding to the vacuum of the anomalous mode $|0\rangle_a$
is shown in panel~(b). Panel~(c) shows the total (i.e. the sum of the
coherent and incoherent parts) density function and panel~(d) is a zoom on
the $x\approx 0$ part of panel~(c) --- note that the density of 
incoherent atoms at the soliton notch is about 14\% of the maximal density
of the surrounding condensate. }
\label{depla}
\end{figure}

   The depletion in the Bogoliubov vacuum of the anomalous mode is not
relevant for dark solitons in current experiments

\begin{itemize}

\item The vacuum $|0\rangle_a$ is not a ground state of the Bogoliubov
Hamiltonian (\ref{HB}). Creation of a quasiparticle in this vacuum with
the creation operator $\hat a^{\dagger}$ actually {\it decreases} the
energy. The Bogoliubov vacuum is {\it not} the state of thermodynamic
equilibrium at $T=0$.

\item For large enough $N$ the density of incoherent atoms in the state
$|0\rangle_a$ exceeds the density of the condensate. The linearized
Bogoliubov theory breaks down and the calculation of depletion in the
state
$|0\rangle_a$ is not self-consistent.

\item The Bogoliubov vacuum $|0\rangle_a$ is {\it not} the state with
minimal possible depletion. When $v(x)\neq 0$ quasiparticles cannot be
identified with incoherent atoms, compare Eq.(\ref{deltaphi}), and the state
with no quasiparticles $|0\rangle_a$ is not the state with minimal number
of incoherent atoms. It is possible to construct a state with a very small
depletion as compared to the depletion in the state $|0\rangle_a$, see the
next Section.

\end{itemize}

\section{ The minimal depletion state }

  Solitons are excited experimentally employing a phase imprinting method
\cite{Hannover}. We assume that before the phase imprinting the system of
$N$ atoms is in its ground state. The ground state is a condensate
(\ref{phi0}) with all Bogoliubov modes in a vacuum state. There are no
anomalous modes and the density of incoherent atoms is very small as
compared to the density of the condensate. The ground state is very close
to a perfect condensate with {\it all} atoms in the condensate wave
function (\ref{phi0}). We approximate the initial condensate by a perfect
condensate without any depletion.

  For a perfect initial condensate the phase imprinting can be thought of
as a nearly instantaneous operation on the condensate wave function. Right
after this operation one gets a condensate with an antisymmetric wave
function $\phi(x)=-\phi(-x)$. To focus on the quantum depletion and not on
classical evolution of $\phi(x)$ we assume that $\phi(x)=\phi_1(x)$. The
initial dark soliton is a (nearly) perfect condensate with the wave
function (\ref{phi1}). In the Hannover experiment \cite{Hannover} they
create moving grey solitons while here, for the sake of simplicity, we
assume a stationary dark soliton. We think that in spite of this,
qualitative comparison with the experiment is still possible.

  The Bogolubov vaccum $|0\rangle_a$ is an eigenstate of the $N$-body
Hamiltonian. For large $N$ the $|0\rangle_a$ has substantial depletion.
The initial state without depletion is far from $|0\rangle_a$ or from any
other eigenstate. The initial state is not stationary, its depletion
may grow from zero to values which far exceed the depletion in the
Bogolubov vacuum. In the following we quantify this effect.

  In the framework of the Bogoliubov theory truncated to the anomalous
mode (\ref{HB}), the initial depletion-free soliton can be very well
approximated by a variational state $|\psi\rangle$ that minimizes the
initial number of atoms depleted from the background condensate
$\phi_1(x)$

\begin{equation}
dN(0)\;=\;
\langle\psi| d\hat N(0) |\psi\rangle \;.
\end{equation}
To find this optimal state it is convenient to introduce bosonic operators
$\hat b$ and $\hat b^{\dagger}$,

\begin{equation}
\hat b\;=\;
 \frac{1}{2}\sqrt{\frac{\Omega}{|\omega|}}
 \left(\hat a + \hat a^\dagger\right)
+\frac{1}{2}\sqrt{\frac{|\omega|}{\Omega}}
 \left(\hat a - \hat a^\dagger\right) \;,
\label{bop}
\end{equation}
where 
\begin{equation}
\Omega\;=\;2^{-\frac{13}{12}}(3g)^{1/3}\;.
\end{equation}
In terms of these new bosonic operators the initial number of depleted
atoms (\ref{dN}) becomes

\begin{equation}
d\hat N(0)~=~
\frac{2\sqrt{2}\Omega}{3}~\hat b^{\dagger}\hat b~+~
\left(
\frac{\sqrt{2}\Omega}{3}-\frac12
\right)~.
\end{equation}
This operator is a sum of a nonnegative operator $\sim\hat b^{\dagger}\hat
b$ and a constant. $\hat b^{\dagger}\hat b\geq 0$ is minimized by the
unique vacuum $|0\rangle_b$ of the bosonic annihilation operator $\hat b$,
$\hat b|0\rangle_b=0$. The minimal depletion state is
$|\psi\rangle=|0\rangle_b$. This vacuum is different from the Bogoliubov
vacuum $|0\rangle_a$,

\begin{equation}
|0\rangle_b \; \neq \; |0\rangle_a \;.
\end{equation}
In this optimal state $|0\rangle_b$ the initial number of depleted atoms
in the soliton notch scales with the number of atoms as $N^{1/3}$. The
initial density of incoherent atoms in the notch

\begin{equation}
\frac{dN_b(0)}{l_0}\propto\;N^{2/3}\;,
\label{db}
\end{equation}
where $dN_b(0)\;=\;_b\langle 0|d\hat N(0)|0\rangle_b$, grows with $N$ in
the same way as the condensate density in Eq.~(\ref{dbec}). 

  In Fig.2 (a,b) we show that for a realistic value of the scattering
length and other parameters the minimal density of incoherent atoms in the
notch is around $1\%$ of the peak density of the surrounding condensate.
Our minimal depletion state is a very accurate approximation to the
initial perfect condensate with a dark soliton. As both coherent and
incoherent densities grow with $N$ in the same way, compare
Eqs.(\ref{dbec},\ref{db}), the quality of this approximation does not
change with increasing $N$. The negligible depletion in the minimal
depletion state justifies {\it a posteriori} our truncation of the theory
to the anomalous mode.

  Moreover, close to the initial minimal depletion state $|0\rangle_b$ we
can safely use the noninteracting Bogoliubov theory. This theory assumes
that the density of incoherent atoms is small as compared to the density
of the condensate. In contrast, for large $N$ the Bogoliubov vacuum
$|0\rangle_a$ does not satisfy this assumption.

\begin{figure}
%\centering \leavevmode \epsfxsize=4cm
\centering
{\epsfig{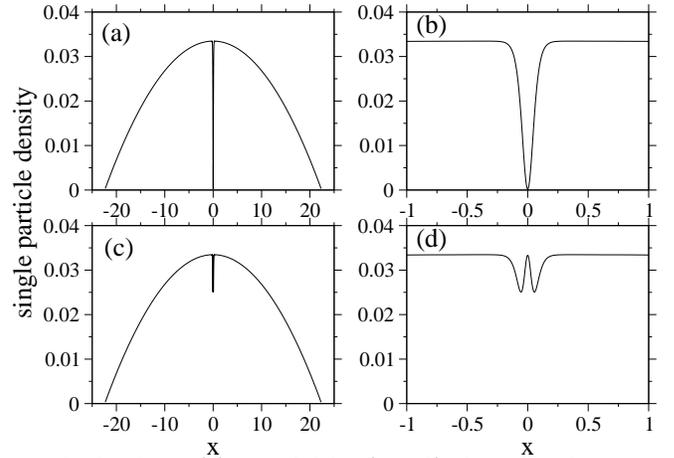}}
\caption{ Panel~(a): the initial ($t=0$) single particle density function
for the system initially in the minimal depletion state $|0\rangle_b$
(the values of the parameters correspond to the Hannover experiment
{\protect \cite{Hannover}}). Panel~(b) is a zoom on the $x\approx 0$
part of panel~(a) --- the density of the incoherent atoms at the soliton
notch is about 1\% of the maximal condensate density. Panels~(c)-(d): the
same as in the previous panels but for
$t_G=0.42\;(\pi/\sqrt{2}\omega_x)=10.6\;$ms. }
\label{deplb}
\end{figure}

\section{ Time evolution of the number of incoherent atoms }

   The initial density of incoherent atoms in the notch 
(calculated in the $|0\rangle_b$ state) is negligible, see
Fig.2. The initial dark soliton is very dark. This small depletion is not
stationary because the initial minimal depletion state $|0\rangle_b$ is
not an eigenstate of the Bogoliubov Hamiltonian (\ref{HB}). From
Eqs.~(\ref{dN},\ref{bop}) one can easily deduce time evolution
of the number of depleted atoms

\begin{eqnarray}
dN_b(t)\;&=&\;
_b\langle 0|d\hat N(t)|0\rangle_b\;=\;
\nonumber\\
&&
dN_b(0)+\frac{(\Omega^2-\omega^2)^2}{3\sqrt{2}\omega^2\Omega}
\sin^2(\omega t)\;.
\label{dN(t)}
\end{eqnarray}
The coefficient in front of the $\sin^2(\omega t)$ function scales with
the total number of atoms as $N^1$ which implies that the maximal density
of depleted atoms in the soliton notch

\begin{equation}
\frac{ dN_b\left( t=\frac{\pi}{2|\omega|} \right) }{ l_0 } \;\propto
\;N^{4/3}\;.
\label{max}
\end{equation}
Hence, for large enough $N$ the density of depleted atoms in the soliton
notch will reach the density of the condensate (\ref{dbec}) before
it reaches the maximum (\ref{max}) at $t=\pi/2|\omega|$.

  Eq.(\ref{dN(t)}) might suggest that the system will go through a series
of periodic revivals of the initial state $|0\rangle_b$ with a period of
$\pi/|\omega|$. This is not a correct conclusion for large $N$. After 
a greying time

\begin{equation} 
t_G<\frac{\pi}{2|\omega|} \;, 
\label{tGl} 
\end{equation}
when the density of incoherent atoms in the notch and the density of the
condensate become comparable for the first time, we cannot use the
noninteracting Bogoliubov theory any more \cite{castin}. After $t_G$ one
can ignore neither anharmonicity in the Hamiltonian (\ref{HB}) of the
anomalous mode nor interaction of the anomalous mode with other Bogoliubov
modes (phonons). We expect that these nonlinearities lead to fast
dephasing and prevent any revivals. The number of depleted atoms in the
notch initially follows Eq.(\ref{dN(t)}) until at around the greying time
$t_G$ their density saturates at a density comparable to the density of
the condensate. This will appear as greying of the dark soliton. The
nonlinear stage deserves more quantitative analysis. However, we expect
that, because of the global instability that shows in the negative
frequency of the anomalous mode, the nonlinearities will not suppress
depletion.

   Eq.(\ref{tGl}) gives a crude upper estimate of $t_G$,
$t_G<\pi/2|\omega|$. Given Eq.(\ref{omega}) we obtain in real time units

\begin{equation}
t_G\;<\;\frac{\pi}{\sqrt{2}\omega_x} \;.
\end{equation}
Here $\omega_x$ is an axial trap frequency. In the Hannover experiment of
Ref.\cite{Hannover}, $\pi/\sqrt{2}\omega_x=25\;$ms while the density of
the incoherent atoms at the soliton notch becomes [according to
Eq.~(\ref{dN(t)})] comparable to the condensate density after $t_G\approx
10\;$ms, see Fig.~\ref{deplb}.

\section{ Discussion and Conclusion }

  It is worthwhile to compare solitons to vortices. There are toy model
calculations in Ref.\cite{vortex} for small $N$ that show some incoherent
atoms in the vortex core. The mechanism is similar to solitons: $2$ atoms
in the state with angular momentum $l=1$ are depleted to states $l=0$ and
$l=2$. $l=0$ gives nonzero density in the core. Depletion to an empty
core/notch is not limited to solitons.

   In contrast to solitons created by phase imprinting, ENS vortices are
created by gradual evaporative cooling to a ground state of a rotating
trap \cite{ENS}. This ground state has no negative frequency anomalous
mode \cite{castin}. This vortex ground state is stationary and has stationary 
depletion in the core. 

  In conclusion, we analyze greying of a dark soliton due to depletion in
the anomalous mode of the Bogoliubov theory. We give approximate analytic
expressions for $u(x)$ and $v(x)$ of the anomalous mode valid in the
Thomas-Fermi regime. These expressions are used to work out formulas for a
time-dependent number of atoms depleted to the notch. We argue that,
contrary to the assumption in Ref.\cite{Law}, the vacuum of the anomalous
mode is far from an optimal approximation to a condensate with a dark
soliton. There is a much better state which minimizes depletion from the
condensate. We argue that this minimal depletion state can serve as a rough
approximation to the state of the system right after creation of the
soliton by the phase imprinting method. The minimal depletion state is not
stationary and the density of incoherent atoms in the notch grows from its
negligible initial value until it fills up the notch. For realistic
parameters like in the Hannover experiment \cite{Hannover} the notch can
be filled up after just $10$ms.

\section*{ Acknowledgments }

  We thank James Anglin for discussions and Diego Dalvit for critical reading
of the manuscript. J.D. was supported in part by NSA. K.S. acknowledges
support of KBN under project 5~P03B~088~21.


\begin{references}
\bibitem{g1d} We estimate the one-dimensional $g$ coefficient by
$4\pi Na_s L/S$ where $a_s$ is the $s$-wave scattering length,
$L=\sqrt{\hbar/m\omega_x}$, and $S$ is an average transverse area
calculated within the Thomas-Fermi approximation.
\end{references}
\end{document}